# Structural, Dielectric, Semiconducting and optical properties of High-Energy Ball Milled YFeO$_3$ Nano-particles

Chandra Bhal Singh[a)], Dinesh Kumar, Narendra Kumar Verma and Akhilesh Kumar Singh[b)]

*School of Materials Science and Technology, Indian Institute of Technology (Banaras Hindu University), Varanasi-221005, India*

Corresponding author: [a)]cbs.sbc@gmail.com, [b)]aksingh.mst@iitbhu.ac.in

**Abstract.** In this work, we report the effects of calcination temperature on structural, dielectric, semiconducting and optical properties of YFeO$_3$ nanoparticles prepared by a high energy ball milling process. The structural analysis of the X-ray diffraction data shows that YFeO$_3$ exists in orthorhombic as well as in hexagonal mixed phase states. The Rietveld analysis confirms that orthorhombic YFeO$_3$ crystallizes into *Pnma* space group. The optical band gap of YFeO$_3$ reduces from 1.96 eV to 1.68 eV with increasing the calcination temperature of the YFeO$_3$ sample. The band gap reducing effect might be attributed to the increased crystallite size and decreased lattice strain which is confirmed by the Williamson-Hall plot method. The obtained low band gap YFeO$_3$ ceramic may provide a new possibility to develop eco-friendly Ferroelectric photovoltaic devices.

## 1. Introduction

Recently, ferroelectric/multiferroic perovskites have attracted for the photovoltaic applications due to high photovoltage and potential for cost reduction of solar panels. Various perovskite ceramics such as BaTiO$_3$, Pb(Zr$_x$Ti$_{1-x}$)O$_3$, BiFeO$_3$ and doped PbTiO$_3$ have shown considerable promise for ferroelectric photovoltaic applications [1-3]. However, there is a critical obstacle of reducing the optical band gap of multiferroic/ferroelectric perovskite ceramics using perovskites free from toxic lead. Various techniques such as doping and strain engineering of ceramic materials are being used to reduce the band gap [4]. Hence, the development of lead-free semiconducting multiferroic/ferroelectric ceramics with low band gap is an emerging research field for photovoltaics. Till date, most of the studied multiferroic/ferroelectric ceramics for photovoltaic applications are either lead-based materials or bismuth-based which are not environment-friendly. In this regard, YFeO$_3$ is a nontoxic and stable semiconducting material for ferroelectric photovoltaic devices. $YFeO_3$ is $RTmO_3$ perovskite-type oxide (with R = rare earth & Tm = transition metal) which shows the semiconducting behaviour at room temperature [5-6]. This ortho-ferrite has multiferroic behaviour and has potential to substitute $BiFeO_3$. It is now a well-known fact that nanomaterials show improved electrical and optical properties. High-energy ball milling is a simple and industrially viable method to produce nano-materials in large scale [7].

In this work, we prepared $YFeO_3$ nanoparticles by the mechanochemical synthesis method using high energy ball milling and investigated their dielectric, semiconducting and optical properties after calcining at various temperatures. The Rietveld refinement of XRD data is used to investigate the crystal structure of $YFeO_3$.

## 2. Experimental Details

The $YFeO_3$ nanoparticles were prepared by mechanochemical synthesis. The synthesis process was carried out using a high energy ball mill (PM400MA, RETSCH, Gmbh, Germany). Stoichiometric weights of constituent oxides, $Y_2O_3$ (Sigma-Aldrich) and $Fe_2O_3$ (Sigma-Aldrich) were used. Both the constituent powders were mixed inside round bottom stainless steel vials with $ZrO_2$ lining using zirconia balls in ethanol medium. After ball milling, the obtained precursors were dried and calcined at $750^{o}C$. The calcination process was carried out at various temperatures in the temperature range $750^{o}C$ to $1200^{o}C$ to get the final phase pure $YFeO_3$. For dielectric measurement, the sintered pellet was electroded using a silver coating. Powder XRD data were collected at room temperature using RIGAKU X-ray diffractometer (MINIFLEX 600) with $Cu_{K\alpha}$ radiations. The particles size and elemental analysis of as prepared $YFeO_3$ nanoparticles were investigated using a scanning electron microscope (ZEISS, Evo Research 18). Temperature-dependent dielectric constant, dielectric loss and resistivity were measured using Keysight-E4990A impedance analyzer. The UV-visible absorbance of $YFeO_3$ nanoparticles was measured using Spectrophotometer (Jasco-722, V-650) in the wavelength range of 200-800 nm. The optical band gap of $YFeO_3$ nanoparticles was calculated by Tauc equation using the absorbance data.

## 3. Results and Discussion

The room temperature XRD patterns in the 2θ range between 10° to 90° for $YFeO_3$ perovskite samples calcined at various temperatures are shown in **Fig. 1**(a). The XRD pattern of the sample calcined at $750^{o}C$ consists of major orthorhombic phase and minor hexagonal phase. The proportion of secondary hexagonal phase of $YFeO_3$ decreases with increasing calcination temperature as reported earlier [8]. As shown in **Fig. 1**(b) by Rietveld fit using FullProf Suite [9], at higher calcination temperature (1200°C) most of the peaks in the XRD pattern can be indexed by using *Pnma* space group of orthorhombic structure. During the Rietveld analysis, we considered that $Y^{3+}$ ions occupy 4c site at $(0.5+\delta x, 0.25, \delta z)$, $Fe^{3+}$ ions occupy 4b site at $(0, 0, 0.5)$, $O^{2-}(1)$ ions occupy 4c site at $(\delta x, 0.25, 0.5+\delta z)$ and $O^{2-}(2)$ ions occupy 4d site at $(0.5+\delta x, \delta y, \delta z)$. **Fig. 1**(b) shows the Rietveld fit between observed and calculated XRD patterns for $YFeO_3$ perovskite sample calcined at 1200°C. The solid scattered data points represent observed XRD pattern while calculated XRD pattern is shown by the plot overlapping to the experimental profile. The lowest plot represents the difference profile between observed and calculated patterns whereas vertical bars represent the position of Bragg's reflections. The refined lattice parameters for $YFeO_3$ samples calcined at various temperatures are given in **Table 1**.

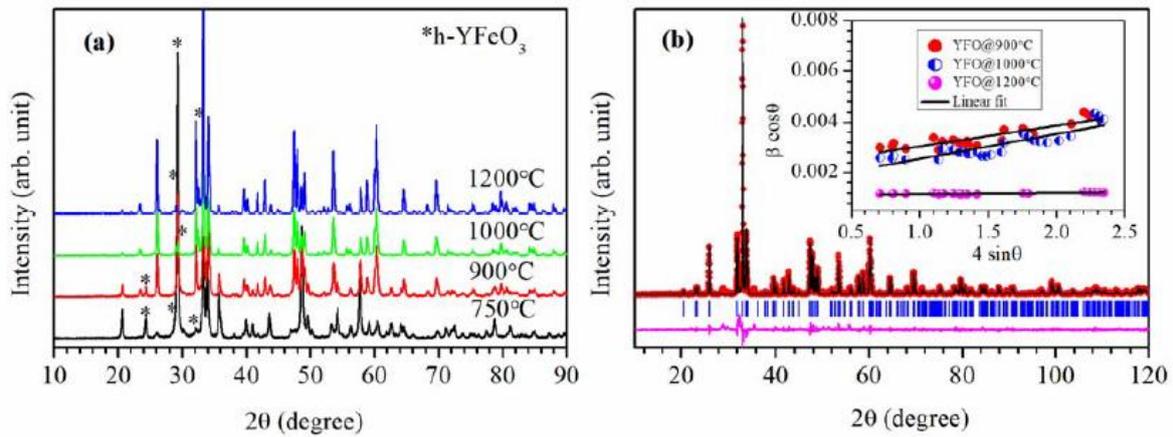

**Figure 1**. (a) Room temperature XRD patterns of $YFeO_3$ (YFO) samples calcined at various temperatures. (b) Rietveld fit for $YFeO_3$ sample calcined at 1200°C. Inset to (b) shows Williamson-Hall plots for $YFeO_3$ samples.

Williamson-Hall (W-H) plot was employed to estimate the average values of crystallite size and lattice strain for $YFeO_3$ samples calcined at various temperatures using the expression given below:

$$\beta \cos\theta = 0.89 \lambda/d + 4\varepsilon \sin\theta$$

where $\beta$ is FWHM of the peaks, $\theta$ is Bragg's angle, $\lambda$ is x-ray wavelength (1.5406 Å), d and $\varepsilon$ are the average value of crystallite size and lattice strain, respectively [10]. The inset of the **Fig. 1**(b) displays W-H plots for $YFeO_3$ samples calcined at various temperatures. The obtained results from W-H plot show that crystallite size of YFO increases from 61.49 nm to 124.65 nm as the calcination temperature increases. The calculated values of the crystallite size, lattice parameters and lattice strain of YFeO$_3$ are also given in **Table 1**.

**Table 1**. Structural parameters (a, b, c & V), crystallite size (d) and lattice strain (ε) for YFeO$_3$ samples obtained from Rietveld refinement and W-H plots.

| Calcination Temp. (°C) | d (nm) | a (Å) | b (Å) | c (Å) | V (Å$^3$) | ε |
|---|---|---|---|---|---|---|
| 900 | 61.49 | 5.5902(5) | 7.6023(7) | 5.2832(6) | 224.53(4) | 8.16×10$^{-4}$ |
| 1000 | 87.89 | 5.5913(4) | 7.6055(5) | 5.2851(4) | 224.75(3) | 9.84×10$^{-4}$ |
| 1200 | 124.65 | 5.5915(1) | 7.6052(2) | 5.2826(1) | 224.64(1) | 4.90×10$^{-5}$ |

**Figs. 2**(a, b, c) show the SEM images of YFO nanoparticles calcined at 750°C, 900°C and 1200°C, respectively. **Fig. 2**(a) is recorded at high resolution. The particle size of YFeO$_3$ nanoparticles calcined at 750°C is around 30-70 nm. The particle size increases with increasing the calcination temperature from 750 to 1200°C. The particle size for the sample calcined at 1200°C is obtained to be 700 nm. Figure 2(d) shows EDS spectrum for YFeO$_3$. The EDS spectrum confirms the presence of Y, Fe and O elements only. No other impurity element is present in the sample.

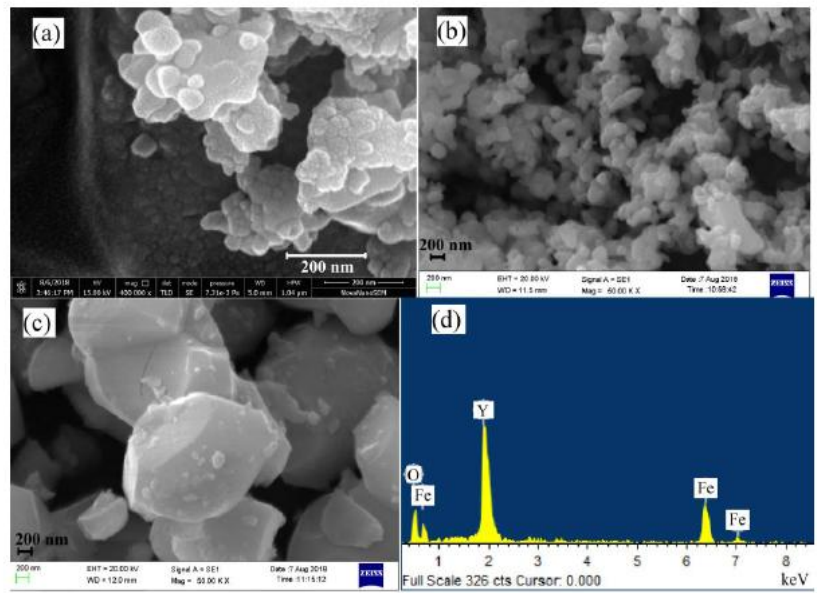

**Figure 2.** SEM images of YFeO$_3$ nanoparticles calcined at (a) 750°C (b) 900°C and (c) 1200°C, (d) Corresponding EDS Spectrum of YFeO$_3$ nanoparticles.

Fig. 3(a) shows the temperature dependence of dielectric permittivity of YFeO$_3$ at frequency 103 Hz, 1990 Hz and 4960 Hz. There are two relaxation peaks, first one below 30°C and a second peak around 230°C. The dielectric permittivity shows a shoulder on lower temperature side and starts decreasing from 30°C to 75°C, which may be due to a low temperature phase transition. There is an increase in dielectric permittivity above 100°C which is due to enhancement in conductivity losses [11]. The low value of dielectric permittivity can be attributed to the small size of particles. The dielectric loss shown in **Fig. 3**(b) initially decreases and become almost constant from 75°C to 175°C but further increases with increasing the temperature. The rapid enhancement in dielectric loss above 250°C may be due to formation of defects [12]. As shown in **Fig. 3**(c), decrease in resistivity above 50°C is observed which confirms that YFeO$_3$ has semiconducting nature. Another peak in resistivity at higher temperature is seen which correspond to dielectric permittivity peak. Optical absorbance spectra and corresponding Tauc plots of YFeO$_3$ powders are shown in **Figs. 3**(d-g). Absorption edges of YFeO$_3$ powders show a red shift with increasing the calcination temperature from 750°C to 1200°C. From Tauc's plot (αhν)$^2$ vs. (hν) the band gap is calculated, where α is linear optical absorption coefficient of YFeO$_3$. The band gap of YFeO$_3$ is found to be 1.96, 1.88 and 1.68 eV for sample calcined at 750°C, 900°C and 1200°C respectively. The decrease in band gap can be correlated to decrease in lattice strain and increase in crystallite size as shown in W-H plot.

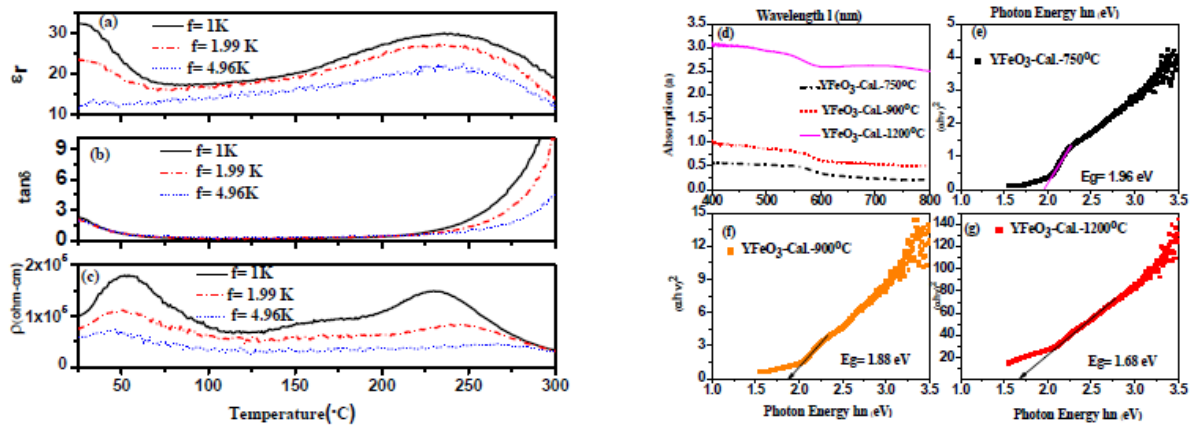

**Figure 3**. (a) Temperature dependent permittivity, (b) dielectric loss and (c) resistivity of YFeO$_3$ sintered at 1300°C, (d) Absorption spectra and optical band gap of YFeO$_3$ powder calcined at (e) 750°C, (f) 900°C and (g) 1200°C.

## 4. Conclusions

In summary, the high energy ball milling process is shown to be efficient in the preparation of YFeO$_3$ nanoparticles. The analysis of the X-ray diffraction data indicates that the structure of YFeO$_3$ consists of coexisting orthorhombic and hexagonal phases. The Rietveld analysis of the X-ray diffraction reveals that orthorhombic phase (*Pnma* space group) is the dominant phase which increases with increasing calcination temperature. The analysis of the HRSEM/SEM micrographs shows that particle size increases with enhancing synthesis temperature. The calcination temperature of the high energy ball milled sample is the key factor in the band gap engineering of YFeO$_3$ multiferroic material. The low band gap and semiconducting behaviour of YFeO$_3$ will be useful to develop and design the absorbing layer of the ferroelectric photovoltaic devices.

**Acknowledgments**

We are thankful to the Central Instrumental Facility (CIF), Indian Institute of Technology (BHU), Varanasi for providing SEM and HRSEM facilities.